\documentclass[a4paper]{jpconf}
\usepackage{graphicx}
\begin{document}
\title{Hot subdwarf binaries -- Masses and nature of their heavy compact companions}

\author{Stephan Geier$^{1}$, Uli Heber$^{1}$, Heinz Edelmann$^{1}$, Thomas Kupfer$^{1}$, Ralf Napiwotzki$^{2}$ and Philipp Podsiadlowski$^{3}$}

\address{$^{1}$ Dr.~Remeis--Sternwarte, Institute for Astronomy, University Erlangen-N\"urnberg, Sternwartstr. 7, 96049 Bamberg, Germany}
\address{$^{2}$ Centre of Astrophysics Research, University of Hertfordshire, College Lane, Hatfield AL10 9AB, UK}
\address{$^{3}$ Department of Astrophysics, University of Oxford, Keble Road, Oxford OX1 3RH, UK}

\ead{geier@sternwarte.uni-erlangen.de}

\begin{abstract}
Neutron stars and stellar-mass black holes are the remnants of massive stars, which ended their lives in supernova explosions. These
exotic objects can only be studied in relatively rare cases. If they are interacting with close companions they become bright X-ray sources. If they are
neutron stars, they may be detected as pulsars. Only a few hundred such systems are presently known in the Galaxy. However, there should be many more
binaries with basically invisible compact objects in non-interacting binaries. 

Here we report the discovery of unseen compact companions to hot subdwarfs in close binary systems. Hot subdwarfs are evolved helium-core-burning stars that have lost most of their hydrogen envelopes, often due to binary interactions. Using high-resolution spectra and assuming tidal synchronisation of the subdwarfs, we were able to constrain the companion masses of 32 binaries. While most hot subdwarf binaries have white-dwarf or late-type main sequence companions, as predicted by binary evolution models, at least $5\,\%$ of the observed subdwarfs must have very massive companions: unusually heavy white dwarfs, neutron stars and, in some cases, even black holes. We present
evolutionary models which show that such binaries can indeed form if the system has evolved through two common-envelope phases. This new connection between hot subdwarfs, which are numerous in the Galaxy, and massive compact objects may lead to a tremendous increase in the number of known neutron stars and black holes and shed some light on
this dark population and its evolutionary link to the X-ray binary population. 

\end{abstract}

\section{Introduction}

Neutron stars and stellar-mass black holes are the remnants of massive stars ending their lifes in supernova explosions. Detecting these exotic objects is possible when they are in a close orbit with another star. If matter is transferred from the companion star to the compact object, bright X-rays are emitted. Without ongoing mass transfer the companion remains invisible, but can be detected indirectly from the reflex motion of the visible star, which causes periodic variations of its radial velocity (RV). These variations are measureable via the Doppler effect from spectral line shifts. Stellar evolution models predict the existence of a hidden population of such compact objects. Subdwarf B or sdB stars are helium core burning stars of about half a solar mass with very thin hydrogen envelopes \cite{heber1}. A large fraction of sdB stars resides in close binaries \cite{maxted2}, \cite{napiwotzki8}.

Because the components' separation in these systems is much less than the size of the subdwarf progenitor in its red-giant phase, these systems must have experienced a common-envelope and spiral-in phase \cite{han1}, \cite{han2}. In such a scenario, two main-sequence stars of different masses evolve in a binary system. If the primary reaches the red-giant phase and fills its Roche lobe, mass is transferred to the companion star. When mass transfer is unstable, a common envelope is formed. Due to friction with the envelope, the two stellar cores spiral towards each other until enough orbital energy has been deposited within the envelope to eject it. The end product is a much closer system containing the core of the giant, which then may become an sdB star, and a main-sequence companion. When the latter reaches the red-giant branch, another common-envelope phase is possible and can lead to a close binary consisting of a white dwarf and an sdB star. In all known cases the companions are either white dwarfs or late-type main-sequence stars.

\section{Analysis method}

\begin{figure}[t!]
\begin{center}
	\resizebox{10cm}{!}{\includegraphics{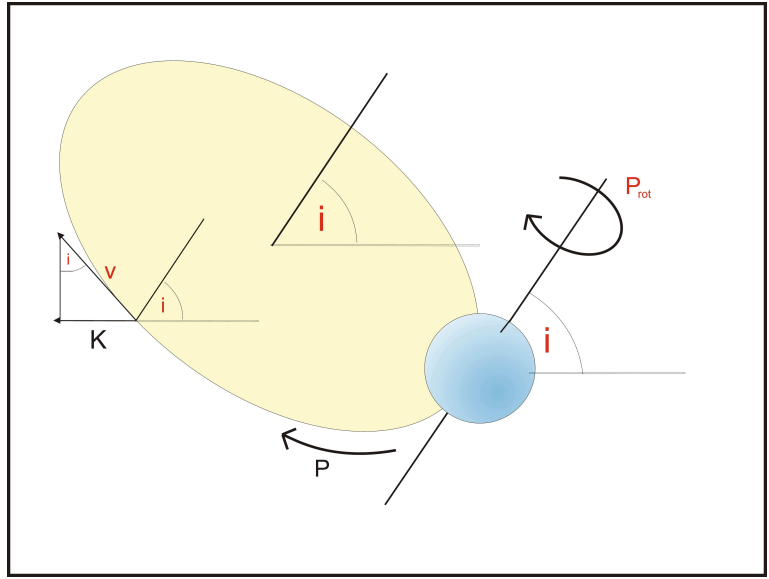}}
	\caption{Schematic view of a single-lined binary system.}
	\label{synchro}
\end{center}
\end{figure}

Since the spectra of close binary subdwarfs are mostly single-lined, they reveal no information about the orbital motion of the sdB stars' companions. However, in very close systems, the rotation of the sdB star is expected to be tidally locked to the orbital motion. This allows us to constrain the inclination angle of the system and to derive the companion mass (Figs.~\ref{synchro}, \ref{synchro2}). In order to derive the parameters, the mass of the sdB primary has to be known. Binary population synthesis models constrain the mass range for the sdB binaries in question. The mass distribution shows a sharp peak at about $0.47\,M_{\rm \odot}$, which we adopted to derive the most probable companion masses.

From the primaries orbital solution only the period $P$ and the projected RV semi-amplitude $K$ can be derived. The mass function \mbox{$f_{\rm m}=\frac{M_{\rm comp}^3 \sin^3i}{(M_{\rm comp}+M_{\rm sdB})^2}=\frac{P K^3}{2 \pi G}$} then  provides a lower limit for the companion mass $M_{\rm comp}$ for a given sdB mass $M_{\rm sdB}$ (Fig.~1). If the primary is synchronised (Fig.~2) the orbital period $P$ equals the rotation period $P_{\rm rot}$. 
\begin{itemize}
\item The stellar radius $R_{\rm sdB}$ is given by the mass radius relation \mbox{$R_{\rm sdB} = \sqrt{M_{\rm sdB}G/g}$}. 
\item The surface gravity $\log{g}$ can be obtained by a quantitative spectral analysis (for details see \cite{geier}). 
\item The rotational velocity \mbox{$v_{\rm rot}=2\pi R_{\rm sdB}/P$} can then be calculated and the projected rotational velocity $v_{\rm rot}\sin{i}$ can be measured from the spectral line broadening of weak metal lines in the case of slow rotators \cite{geier2} and of Balmer and helium lines for fast rotators \cite{geier}. 
\end{itemize}
Now the inclination angle $i$ can easily be derived and the mass function can be solved for reasonable values of $M_{\rm sdB}$.

The stars in our sample were observed with the high-resolution ($R=20\,000-48\,000$) spectrographs UVES at the ESO\,VLT, HIRES at the Keck telescope, HRS at the Hobby Eberly Telescope (HET), FEROS at the ESO\,2.2\,m telescope and FOCES at the CAHA\,1.5\,m telescope.

\begin{figure}[t!]
\begin{center}
	\resizebox{10cm}{!}{\includegraphics{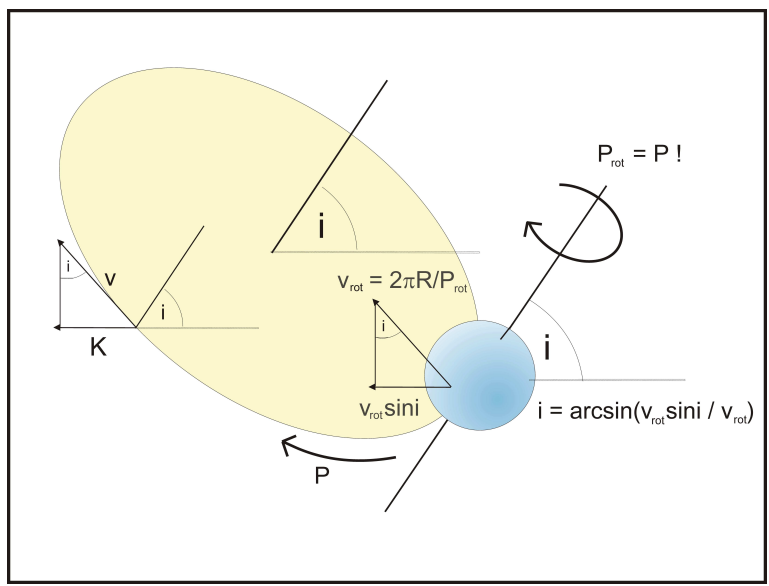}}
	\caption{Schematic view of a single-lined binary system with synchronised rotation.}
	\label{synchro2}
\end{center}
\end{figure}

\section{Nature of the unseen companions}

\begin{figure}
\begin{center}
	\resizebox{8cm}{!}{\includegraphics{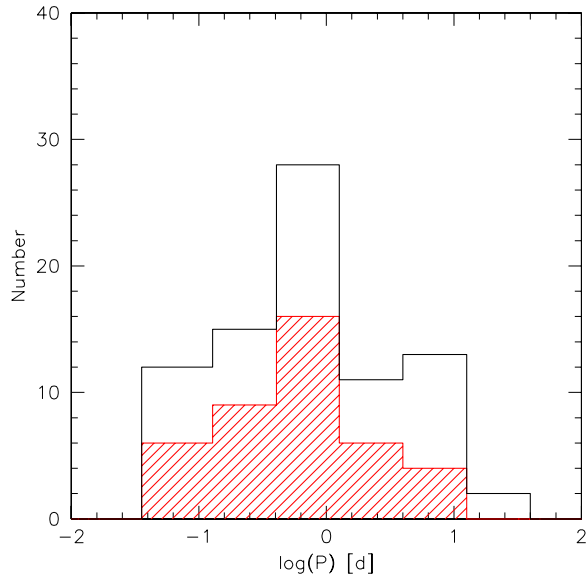}}
	\caption{Number of binaries plotted against the logarithm of their orbital periods. The solid blank histogram marks all known sdB binaries from the catalogue of Ritter \& Kolb \cite{ritter}. The shaded histogram marks the studied sample of 41 binaries with a focus on systems with rather short periods.}
	\label{periods}
\end{center}
\end{figure}

Today, 81 sdB stars in close binary systems have been studied and their orbital parameters have been derived (see the catalogue of Ritter \& Kolb \cite{ritter}). From 51 radial velocity variable sdBs in our sample, we selected 41 binaries with known orbital parameters or half of the known sample (see Fig.~\ref{periods}). From these, 32 could be solved consistently under the assumption of synchronisation. In 9 cases the sdB primaries spin faster than synchronised. There are no spectral signatures of companions visible. Main sequence stars with masses higher than $0.45\,M_{\rm \odot}$ could therefore be excluded because of their high luminosities in comparison to the sdB stars. In this case spectral features of the cool secondary (e.g. Mg\,{\sc i} lines at $\approx5\,160\,{\rm \AA}$) get visible in the spectra and a flux excess in the infrared appears, which can be detected using 2MASS photometry \cite{lisker}. 

Another possibility to detect M dwarf companions are reflection effects in the binary light curves. Some of our programme stars have already been checked for modulations in their light curves. Unfortunately, this method only works, if the binary inclination is high and the orbital period short enough. Synthetic light curve modelling of sdB+M systems shows that the expected amplitude of a reflection effect drops below $1\,{\rm mmag}$ for orbital periods longer than $0.5\,{\rm d}$. This period therefore provides an upper limit for the detectability of reflection effects in sdB binaries from the ground.

\begin{itemize}
\item In seven sdB binaries the companion has been identified as late M star. In addition to the derived companion masses, reflection effects are visible in their lightcurves. 
\item Nine companions have to be white dwarfs, because of the derived masses and the absence of reflection in their lightcurves or flux contributions in their spectra. One sdB has a massive white dwarf companion and short orbital period. It qualifies as candidate for SN\,Ia.
\item In seven cases this distinction cannot be made. The companions may be either late M stars or white dwarfs. 
\item Eight systems have massive compact companions near or even exceeding the Chandrasekhar limit. Their companions therefore may be either unusually heavy white dwarfs, neutron stars or black holes (see Fig.~3).
\end{itemize}

\begin{figure}
\begin{center}
	\resizebox{14cm}{!}{\includegraphics{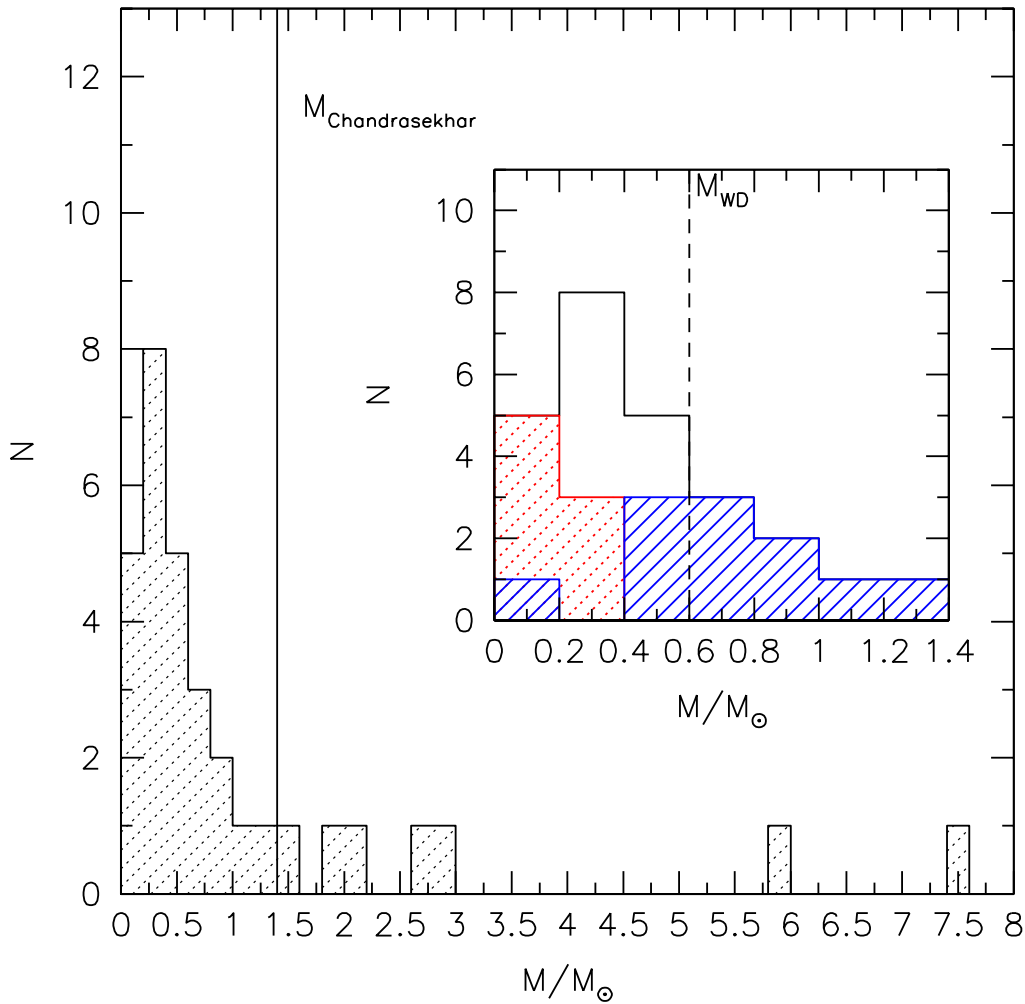}}
	\caption{Mass distribution of the unseen companion stars. The companion mass is plotted against the total number of binaries under the assumption of canonical sdB mass $0.47\,M_{\rm \odot}$. The solid vertical line marks the Chandrasekhar limit. The last bin at $7.5\,M_{\rm \odot}$ is a lower limit. A detail of the mass distribution is shown in the inlet. The shaded solid histogram shows the fraction of subdwarfs with confirmed compact companions, the shaded dashed one the detected M dwarf companions. The dashed vertical line marks the average WD mass.}
	\label{synchro2}
\end{center}
\end{figure}

\section{Orbital synchronisation of sdB binaries}

The assumption of orbital synchronisation in close binary sdBs is essential for our analysis. Theoretical models predict synchronisation to be established up to orbital periods of about half a day to two days \cite{zahn2}, \cite{tassoul}. But it has to be pointed out that tidal dissipation in radiative stellar envelopes is still poorly understood and the timescales of synchronisation differ by orders of magnitude depending on the models. Empirical evidence is therefore needed to constrain such models and to answer the question which close binary sdBs are synchronised.

The timescale of the synchronisation process is highly dependent on the tidal force exerted by the companion. If the companion is very close and the orbital period therefore very short, synchronisation is established much faster than in binaries with longer orbital periods. If the sdB in a binary with given orbital period is proven to be synchronised, all other sdBs in binaries with shorter orbital periods should be synchronised as well. Although the timescales also scale with sdB radius and companion mass, the orbital period is dominating at first order. Most of the non-synchronised systems in our sample have orbital periods exceeding $1.3\,{\rm d}$.

\begin{figure}[t!]
\begin{center}
        \resizebox{7.5cm}{!}{\includegraphics{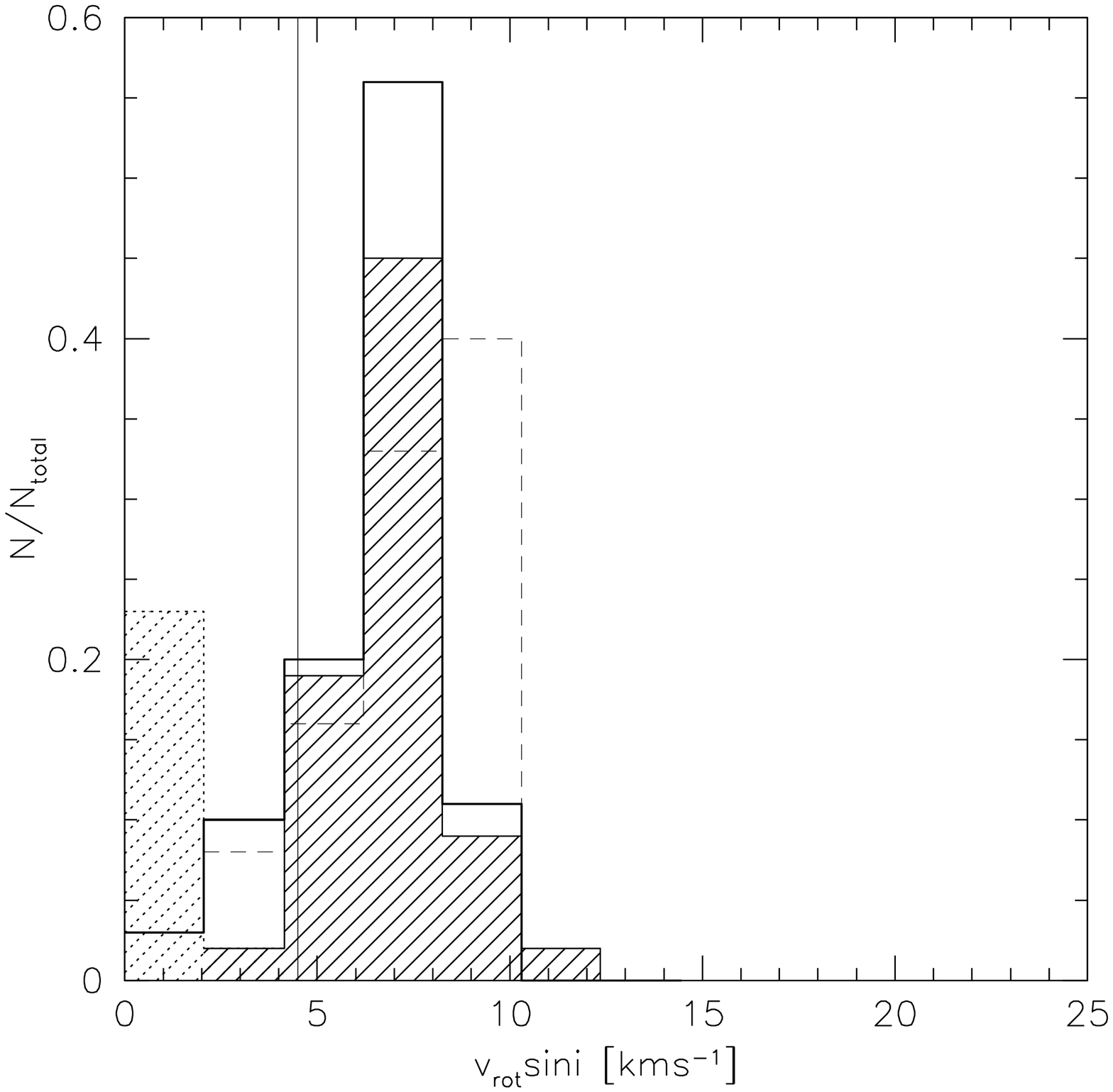}}
	\resizebox{7.5cm}{!}{\includegraphics{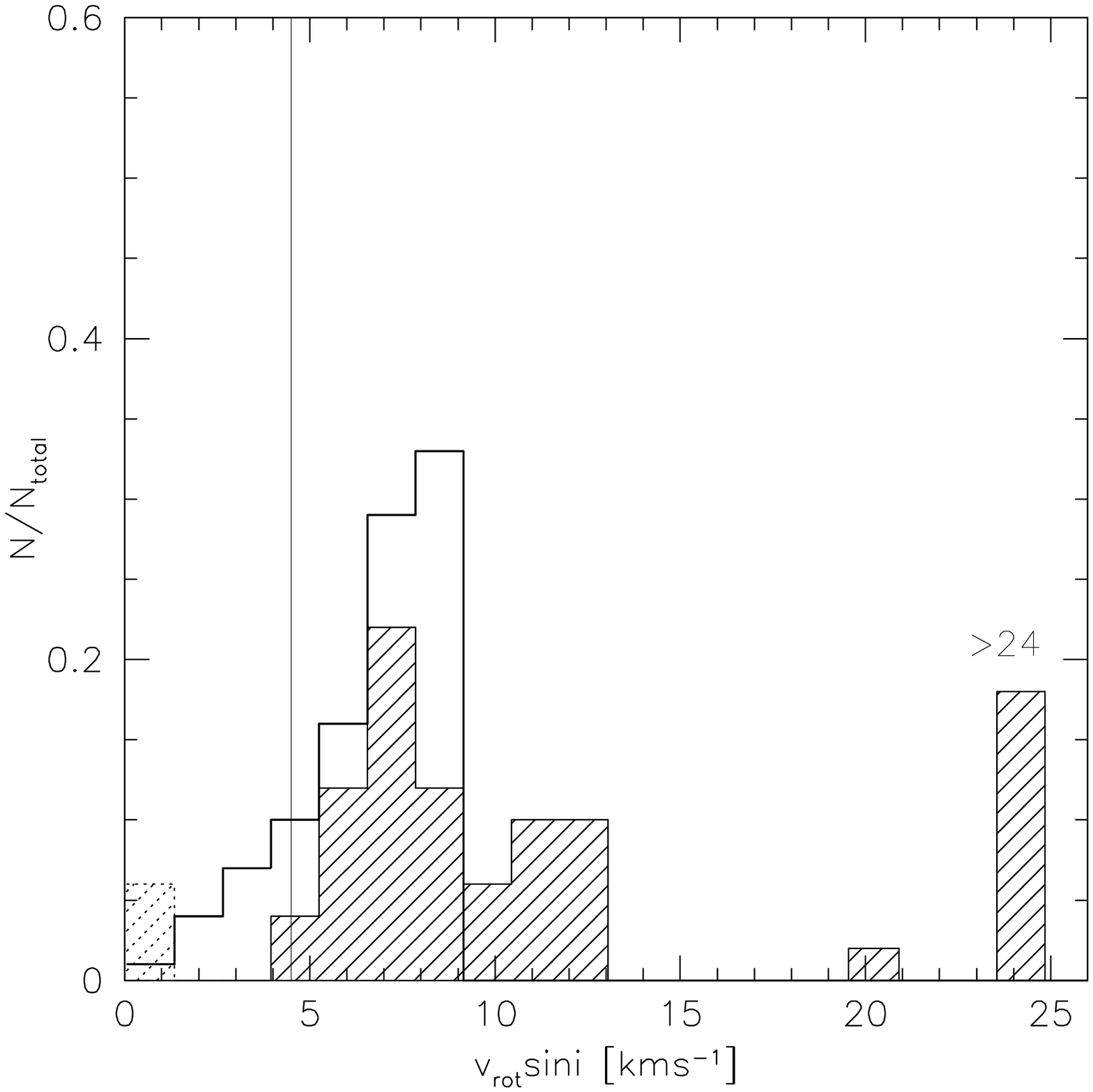}}
	\caption{{\it Left panel} The measured $v_{\rm rot}\sin{i}$ of 49 single sdBs is plotted against relative fraction of stars as shaded histogram. The size fo the bins is given by the average error of the measurements. The blank histogram marks the expected uniform distribution of $v_{\rm rot}\sin{i}$ under the assumption of randomly oriented polar axes and the same rotational velocity $v_{\rm rot}=8.3\,{\rm kms^{-1}}$ for all stars. The dashed histogram shows this distribution for $v_{\rm rot}=9.0\,{\rm kms^{-1}}$. The solid vertical line at $v_{\rm rot}\sin{i}\approx4.5\,{\rm kms^{-1}}$ marks the detection limit. All sdBs with lower $v_{\rm rot}\sin{i}$ are stacked into the first bin (dotted histogram). {\it Right panel} The measured $v_{\rm rot}\sin{i}$ of 51 RV variable sdBs is plotted in the same way. The size of the bins is given by the average error of the measurements and is therefore slightly different than in the left panel. The blank histogram marks the uniform distribution for $v_{\rm rot}=8.3\,{\rm kms^{-1}}$. All sdBs with $v_{\rm rot}\sin{i}$ higher than $24\,{\rm kms^{-1}}$ are stacked into the last bin.}
	\label{vrotdistrib_RV}
\end{center}
\end{figure}

To study the influence of close companions on the rotational properties of sdBs we measured the projected rotational velocities of 49 single sdBs from the SPY survey \cite{lisker}. We found that single sdBs are slow rotators spinning with an almost uniform rotational velocity of about $8\,{\rm kms^{-1}}$. A comparison with the $v_{\rm rot}\sin{i}$-distribution of our close binary sample revealed that the rotation of sdBs in close binary systems is clearly affected by the tidal influence of the invisible companions (see Fig.~\ref{vrotdistrib_RV}). This alone does not prove that all sdBs in these binaries are synchronised, but it shows that the tidal influence of the companions is strong enough to significantly change the rotational properties of the sdBs. Although these stars may be not yet be synchronised, the synchronisation process is already at work.

The parameters of the eclipsing binaries PG\,1336$-$018 \cite{vuckovic2}, HS\,0705$+$6700 \cite{drechsel} and HW\,Vir \cite{edelmann3} derived from light curve analyses are consistent with the parameters derived with our method assuming synchronised orbits. This essentially means that the calculated $v_{\rm rot}\sin{i}$ for synchronous rotation is consistent with the measured value. In eclipsing systems, all these parameters can be measured. This provides clear empirical evidence that at least the upper layers of the stellar envelope are synchronised to the orbital motion of the eclipsing sdB binaries in our sample. We therefore conclude that all sdBs in close binaries with orbital periods up to $0.12\,{\rm d}$ should be synchronised as well.

Two known sdBs clearly show ellipsoidal variations in their light curves with periods exactly half the orbital periods (KPD\,0422+5421 \cite{orosz}; KPD\,1930+2752 \cite{maxted}, \cite{geier}). This alone is only an indication for tidal synchronisation. To really prove synchronisation it is necessary that the stellar parameters determined independently from the light curve analysis are consistent with a synchronised orbit. This is the case for KPD\,0422+5421 as well as KPD\,1930+2752. Both ellipsoidal variable systems have very short periods of $\approx0.1\,{\rm d}$ and high inclinations. Otherwise ellipsoidal variations are very hard to detect. 

This was possible in the case of the sdB+WD binary PG~0101$+$039 using a high precision light curve obtained with the MOST satellite \cite{geier2}. We found a strong indication that the surface rotation of the sdB star PG~0101$+$039 is tidally locked to its orbit. Hence, other sdB stars in close binaries should also be synchronised if their orbital period is less than that of PG~0101$+$039 (P\,=\,$0.567\,{\rm d}$). We conclude that tidally locked surface rotation is at least established in sdB binaries with orbital periods of less than half a day.

An independent method to proof orbital synchronisation is provided by asteroseismology. Van Grootel et al. were able to reproduce the main pulsation modes of the short period pulsating sdB Feige 48 ($P\approx0.38\,{\rm d}$), derived the surface rotation from the splitting of the modes and concluded that the subdwarf rotates synchronously \cite{vangrootel}. Charpinet et al. reach a similar conclusion for the short period eclipsing binary PG\,1336$-$018 ($P\approx0.10\,{\rm d}$) \cite{charpinet5}. Asteroseismic analyses revealed that sdB binaries up to orbital periods of about $0.4\,{\rm d}$ are synchronised. We therefore conclude that all sdBs in close binaries with shorter periods should be synchronised as well.

From the eight candidate sdB+NS/BH binaries found in our sample four have orbital periods shorter than $0.4\,{\rm d}$ where synchronisation could be proven by asteroseismology. Another two binaries have periods shorter than $0.6\,{\rm d}$, the lower limit for synchronised rotation derived by binary light curve analysis. And the longest binary period is only $0.8\,{\rm d}$. We therefore conclude that all these binaries should be synchronised and the derived parameters are therefore correct.

\section{Formation of sdB+NS/BH binaries}

The evolution that leads to such systems requires an initial binary, consisting of a primary star massive enough to produce a neutron star or black hole, and a companion of typically several solar masses. These systems experience two mass-transfer phases and one supernova explosion (see Fig.~4). The second mass-transfer phase had to be unstable, leading to a common-envelope and spiral-in phase. Population synthesis estimates suggest that of order 1\,\% of all hot subdwarfs should have neutron-star or black-hole companions \cite{podsi}, \cite{pfahl}.

\section{The fraction of NS/BH companions and selection effects}

This high fraction of massive compact companions ($20\,\%$) is in part caused by an interplay of several different selection effects (e.g. RV variability selection, bias on short period binaries, see Fig.~\ref{periods}). Especially the fact that most of the binaries with the highest derived companion masses have low inclination seems suspicious. It can be explained by the limitations of our analysis method. Nevertheless, high inclination systems should be more numerous than binaries seen at low inclination. The fact that no such system was found up to now (81 binaries with solved orbits are known), puts an upper limit of a few percent to the number fraction of sdB+NS/BH binaries.

Both observational constraints and binary population synthesis models suggest a fraction of a few percent for sdBs with NS/BH companions. The fraction we found in our sample is too high to be consistent with these results even if selection effects are taken into account. It has to be pointed out that all of our candidate binaries have short orbital periods ($0.2-0.8\,{\rm d}$) and should therefore be most likely synchronised. Adopting the lowest possible mass for the sdBs ($0.3\,M_{\rm \odot}$, \cite{han1}, \cite{han2}) only two of our candidate binaries ($\approx5\,\%$) have companion masses exceeding the Chandrasekhar limit. This number is consistent with theoretical and observational constraints. We therefore conclude that most sdBs in our candidate systems should have masses lower than canonical.

\begin{figure}
%\begin{center}
	\resizebox{15cm}{!}{\includegraphics{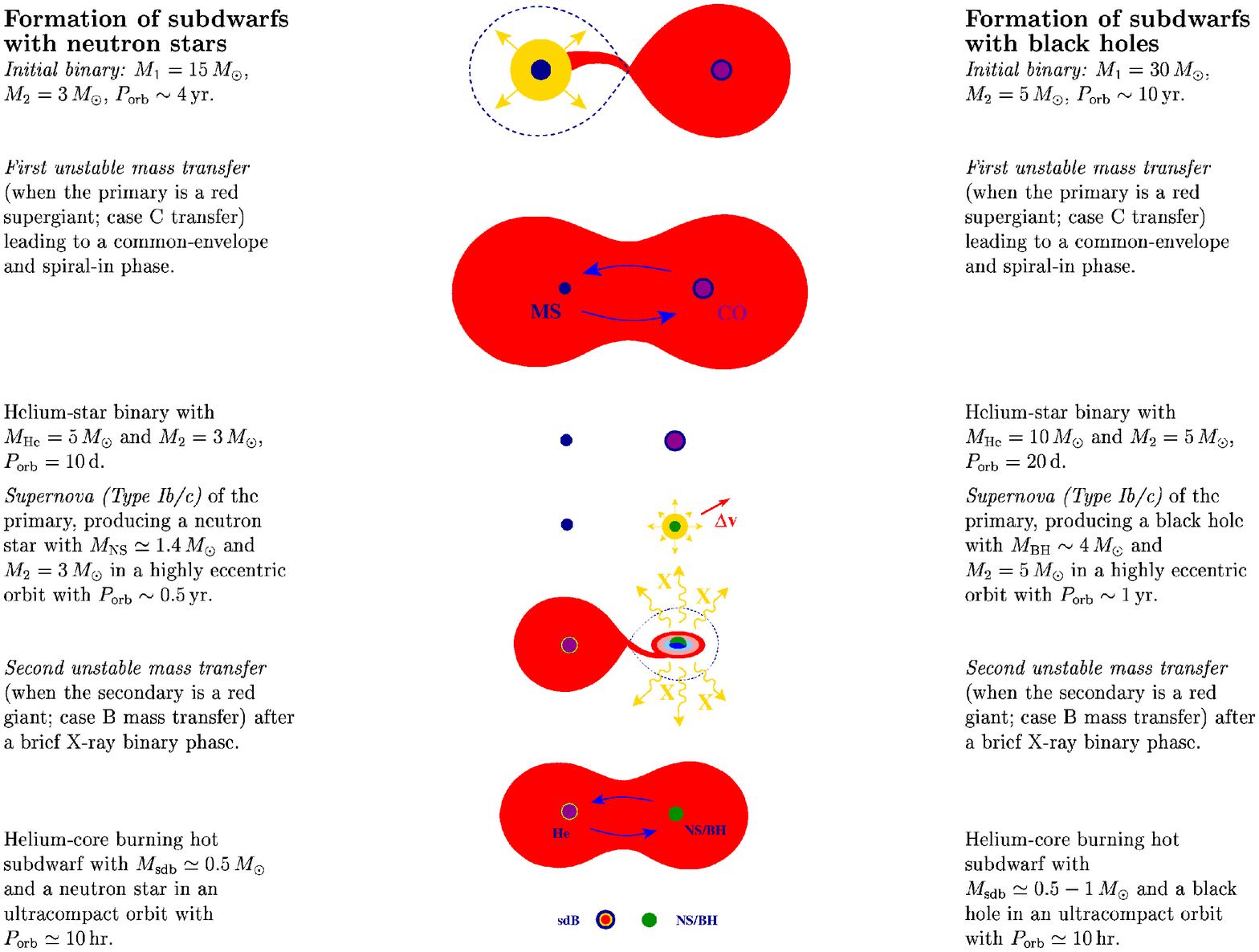}}
	\caption{Schematic diagram of formation scenarios leading to hot subdwarf binaries with neutron-star (right hand panel) or black-hole (left hand panel) companions.}
	\label{heavy}
%\end{center}
\end{figure}

\section{HYPERMUCHFUSS survey}

The results presented here provide strong indications for the existence of sdB+NS/BH binaries, but rely on the assumption of orbital synchronisation and are seriously affected by selection effects. Most of the binaries with massive unseen companion are seen at low inclination. High inclination systems must exist as well and should be even more numerous. In this case a determination of the orbital parameters would be sufficient to prove their existence, because in binaries with such high RV amplitudes the lower limit for the companion mass (see Fig.~1) should already exceed the Chandrasekhar mass. The HYPERMUCHFUSS-survey has been launched in 2008 to search for sdBs with high RV variability and massive compact companions as well as the recently discovered class of hypervelocity stars (see Tillich et al. these proceedings).


\section*{References}

\begin{thebibliography}{19}

\bibitem{heber1} Heber, U. 1986 {\it A\&A} {\bf 155} 33--45

\bibitem{maxted2} Maxted, P. F. L., Heber, U., Marsh, T. R., \& North, R. C. 2001 {\it MNRAS} {\bf 326} 1391--1402 

\bibitem{napiwotzki8} Napiwotzki, R., Karl, C. A., Lisker, T., et al. 2004 {\it Ap\&SS} {\bf 291} 321--324

\bibitem{han1} Han, Z., Podsiadlowski, P., Maxted, P. F. L., Marsh, T. R., \& Ivanova, N. 2002 {\it MNRAS} {\bf 336} 449--466

\bibitem{han2} Han, Z., Podsiadlowski, P., Maxted, P. F. L., \& Marsh, T. R. 2003 {\it MNRAS} {\bf 341} 669--691

\bibitem{geier} Geier, S., Nesslinger, S., Heber, U., et al. 2007 {\it A\&A} {\bf 464} 299--307

\bibitem{geier2} Geier, S., Nesslinger, S., Heber, U., et al. 2008 {\it A\&A} {\bf 477} L13--L16

\bibitem{ritter} Ritter, H., \& Kolb, U. 1998 {\it A\&AS} {\bf 129} 83--85

\bibitem{lisker} Lisker, T., Heber, U., Napiwotzki, R., et al. 2005 {\it A\&A}, 430, 223--243

\bibitem{zahn2} Zahn, J.-P. 1977 {\it A\&A} {\bf 57} 383--394

\bibitem{tassoul} Tassoul, J.-L., \& Tassoul, M. 1992 {\it ApJ} {\bf 395} 259--267

\bibitem{vuckovic2} Vu\v ckovi\'c, M., Aerts, C., \O stensen, R., et al. 2007 {\it A\&A} {\bf 471} 605--615 

\bibitem{drechsel} Drechsel, H., Heber, U., Napiwotzki, R., et al. 2001 {\it A\&A} {\bf 379} 893--904

\bibitem{edelmann3} Edelmann, H. 2008 {\it ASPCS} {\bf 392} 187--194

\bibitem{orosz} Orosz, J. A., \& Wade, R. A. 1999 {\it MNRAS} {\bf 310} 773--783

\bibitem{maxted} Maxted, P. F. L., Marsh, T. R., \& North, R. C. 2000 {\it MNRAS} {\bf 317} L41--L44

\bibitem{vangrootel} van Grootel, V., Charpinet, S., Fontaine, G., \& Brassard, P. 2008 {\it A\&A} {\bf 483} 875

\bibitem{charpinet5} Charpinet, S., van Grootel, V., Reese, D., et al. 2008 {\it A\&A} {\bf 489} 377

\bibitem{pfahl} Pfahl, E., Rappaport, S., \& Podsiadlowski, Ph. 2003, {\it ApJ} {\bf 597} 1036--1048

\bibitem{podsi} Podsiadlowski, Ph., Rappaport, S., \& Pfahl, E. D. 2002 {\it ApJ} {\bf 565} 1107--1133

\end{thebibliography}
\end{document}